\documentclass[a4paper,11pt]{article}
\pdfoutput=1 

\usepackage{amsthm,booktabs,integrable,jheparxiv,xparse,stmaryrd,twistor}
\DeclareUnicodeCharacter{039B}{\ensuremath{\Lambda}}

\usepackage{tikz}
\usepackage{amsthm}

\newtheorem{theorem}{Theorem}
\usetikzlibrary{decorations.markings}

\title{The Conformal $w$-Algebra}

\author[a,b]{Simon Heuveline,}
\author[a]{Nia Robles Del Pino,}
\author[a,b,c]{Andrew Strominger}

\affiliation[a]{Center for the Fundamental Laws of Nature,\\ Harvard University, Cambridge, United States \vspace{0.1cm}}
\emailAdd{simonheuveline@fas.harvard.edu}
\emailAdd{niaroblesdelpino@g.harvard.edu}
\emailAdd{strominger@fas.harvard.edu}

\affiliation[b]{ Black Hole Initiative,\\ Harvard University, Cambridge, United States\vspace{0.1cm}}

\affiliation[c] {OpenAI, San Francisco, United States }

\def\ed{\end{document}}

\def\cw{${\cal L}w_{1+\infty}$}
\def\dw{${\cal L}_\Lambda w_{1+\infty}$}
\def\Cw{${\cal C}w_{1+\infty}$}
\begin{document}

\abstract{  Einstein gravity in four asymptotically flat dimensions is governed by a tower of soft theorems obeying  an $\mathcal{L}w_{1+\infty}$ commutator algebra. This algebra is deformed to 
$\mathcal{L}_\Lambda w_{1+\infty}$ in the conformally related AdS$_4$ and dS$_4$ spacetimes.  Here we construct the  larger ``conformal $w$-algebra", denoted 
$\mathcal{C}w_{1+\infty}$, which acts on conformal gravity in all of these spacetimes. It includes 
$\mathcal{L}w_{1+\infty}$, 
$\mathcal{L}_\Lambda w_{1+\infty}$ and the $\mathfrak{so}(4,2)$ 4D conformal algebra as subalgebras. We show that 
$\mathcal{C}w_{1+\infty}$ is the conformal completion of either 
$\mathcal{L}w_{1+\infty}$ or 
$\mathcal{L}_\Lambda w_{1+\infty}$.   A derivation of 
$\mathcal{C}w_{1+\infty}$ is given as the symmetry of the twistor action of self-dual conformal gravity.}

\maketitle

\flushbottom

\section{Introduction}

Soft algebras, namely \cw\ \cite{Strominger:2021mtt,Guevara:2021abz} for Einstein gravity in flat space and \dw\ \cite{Taylor:2023ajd, Bittleston:2024rqe, Heuveline:2025nmb} in the presence of a cosmological constant, play a central role in controlling the dynamics of classical and quantum Einstein gravity. These theories are a subset of the larger Weyl$^2$ conformal gravity theory, in the sense that all solutions of the Einstein equations, with or without a cosmological constant, are also solutions of the conformal gravity equations. Conformal gravity is also of great interest in its own right, despite its unphysical ghost poles, as an example of a finite quantum theory of gravity when suitably supersymmetrized \cite{Howe:1983sr, Fradkin:1985am, Strominger:1984ed} and as a natural output of the Berkovits-Witten string \cite{Berkovits:2004jj, Berkovits:2004hg, Witten:2003nn}.

In this paper, we extend the soft symmetry analysis to conformal gravity and find the ``conformal $w$-algebra''. It is characterized by a global $\mathfrak{so}(4,2)$ subalgebra, which generates four-dimensional conformal transformations. In section 4, we show that, as expected from the inclusion of the solution spaces, this larger algebra contains both \cw\ and \dw\ as subalgebras. A third conformally flat spacetime is the Einstein cylinder $S^3\times\mathbb{R}$; we conjecture a subalgebra of \Cw\ that describes Einstein gravity on this cosmological spacetime.

These $w$-algebras have a gauge-theory cousin known as the $S$-algebra \cite{Strominger:2021mtt, Guevara:2021abz}. In section 5, we show how \Cw\ acts on the $S$-algebra.

In section 6, we turn the construction around and derive \Cw\ from the action of $\mathfrak{so}(4,2)$ on \cw\ or \dw. This action can be determined by realizing the generators of these algebras as metric modes solving the linearized Einstein equation.  Since Einstein gravity is not conformally invariant, it does not lead to new Einstein modes or to generators of the original algebra. In particular, we find that the special conformal symmetry generators do not map $\mathcal{L}w_{1+\infty}$ to itself. Rather, they produce new linearized  solutions of the conformal gravity equations corresponding  to generators of a much larger algebra. This larger algebra can be constructed by acting in all possible ways and then taking all possible commutators of the generators so obtained.  We show that this procedure indeed leads to the \Cw\ algebra. In other words, the \Cw\ algebra can be thought of as the conformal completion of either the \cw\ or the \dw\ algebra.

Finally, in section 7, we turn to twistor theory, where these observations have a natural home.  We show that the self-dual sector of conformal gravity has a natural twistor description in which \Cw\ arises as the gauge symmetry of the twistor action. Indeed, the existence of the algebra \Cw\ is implicit  in the twistor literature \cite{Penrose:1976js, Adamo:2021lrv, Bittleston:2025jmk, Adamo:2013tja, Costello:2022wso, Bittleston:2022jeq, Costello:2023hmi}; here we hope to make it more explicit and connect to recent developments in gravitational soft algebras.

\newpage
\section{Summary of Conformal Gravity}
\label{sec:confGrav}
The action for 4D conformal gravity is given by
\begin{equation}\label{eq:confgravaction}
S[g] =\frac{1}{\lambda^2} \int_M \, C^{\mu\nu\rho\sigma} C_{\mu\nu\rho\sigma} \,\sqrt{ |g|}\,\dif^4 x \,,
\end{equation}
where $C_{\mu\nu\rho\sigma}$ is the Weyl curvature tensor of a four-dimensional metric $g$ and $\lambda^2$ is a dimensionless coupling constant.
In four dimensions, the Weyl tensor has ten components that decompose into five self-dual and five anti-self-dual components. In spinor notation, this corresponds to the decomposition
\be \label{eq;sdasdWeyl}
C_{\alpha\dot{\alpha}  \beta\dot{\beta} \gamma  \dot{\gamma} \delta \dot{\delta}}=\Psi_{\alpha \beta \gamma \delta}\varepsilon_{\dot{\alpha} \dot{\beta}}\varepsilon_{ \dot{\gamma} \dot{\delta}}+\widetilde{\Psi}_{\dot{\alpha} \dot{\beta} \dot{\gamma} \dot{\delta}}\varepsilon_{{\alpha} {\beta}}\varepsilon_{ {\gamma} {\delta}}\,,
\ee
Here, $\Psi_{\alpha \beta \gamma \delta}=\Psi_{(\alpha \beta \gamma \delta)}$ and $\widetilde{\Psi}_{\dot{\alpha} \dot{\beta} \dot{\gamma} \dot{\delta}}=\widetilde{\Psi}_{(\dot{\alpha} \dot{\beta} \dot{\gamma} \dot{\delta})}$ are totally symmetric spinors.
Using \eqref{eq;sdasdWeyl},
the action \eqref{eq:confgravaction} can be written as
\begin{equation}
S[g] = 
\frac{1}{\lambda^2}\int_M \, 
\left(\Psi^{\alpha \beta \gamma \delta}\Psi_{\alpha \beta \gamma \delta} + \widetilde{\Psi}^{\dot{\alpha} \dot{\beta} \dot{\gamma} \dot{\delta}} \widetilde{\Psi}_{\dot{\alpha} \dot{\beta} \dot{\gamma} \dot{\delta}} \right)\,\sqrt{ |g|}\,\dif^4 x.
\end{equation}
This action can easily be seen to be conformally invariant. It therefore does not depend on the representative $g\in[g]$ of the conformal class $[g]$. The equations of motion are given by the vanishing of the conformally covariant \emph{Bach tensor} $B_{\mu\nu}$ \cite{Bach:1921zdq},
\be
\label{eq:BachFlat}
B_{\mu\nu}\equiv 
 \big( 2\nabla^\rho \nabla^\sigma 
+  R^{\rho\sigma} \big)
C_{\mu\rho\nu\sigma}=0~.
\ee

In the following, we are also interested in the self-dual case. We introduce a totally symmetric Lagrange multiplier field $b^{\alpha \beta \gamma \delta}=b^{(\alpha \beta \gamma \delta)}$ and define the following action
for \emph{self-dual conformal gravity}:
\be
\label{eq:SDCG}
S[g,b]=\int_M \,b^{\alpha \beta \gamma \delta}\Psi_{\alpha \beta \gamma \delta} \,\sqrt{ |g|}\,\dif^4 x\,.
\ee
$S[g,b]$ is simply a Lagrange multiplier action imposing 
\be
\Psi_{\alpha \beta \gamma \delta}=0\,,
\ee
which in particular implies vanishing of the Bach tensor. To see this, use
\begin{equation}
B_{\alpha\beta\dot{\alpha}\dot{\beta}}
= \nabla^{\gamma}{}_{\dot{\alpha}} \nabla^{\delta}{}_{\dot{\beta}} \, \Psi_{\alpha\beta\gamma\delta}
+ \Phi^{\gamma\delta}{}_{\dot{\alpha}\dot{\beta}} \, \Psi_{\alpha\beta\gamma\delta}=0\,,
\end{equation}
where $\Phi_{\alpha \beta \dot{\alpha}\dot{\beta}}$ is the trace-free Ricci tensor in spinor notation. Much of what we discuss pertains to the self-dual, conformally invariant theory \eqref{eq:SDCG}.

Let us review the linearized spectrum of conformal gravity. The most important point for us is that it contains six states, two pairs of spin $2$ states with helicities $\pm2$ and a pair of spin $1$ states with helicities $\pm1$ \cite{Stelle:1977ry, Riegert:1984hf}. The (anti-)self-dual subsector only contains half of these states, those with positive (respectively negative) helicities. Let us describe this in more detail \cite{Adamo:2013tja, Adamo:2018srx}.

In order to obtain the linearized spectrum, we perturb the metric in \eqref{eq:BachFlat} around a flat background
$g_{\mu\nu}=\eta_{\mu\nu}+\varepsilon h_{\mu\nu}$. At first order in $\varepsilon$, \eqref{eq:BachFlat} becomes the fourth-order equation
\bea \label{eq:linearizedBach}
\Box^2h_{\mu\nu}+\partial_\mu V_\nu+\partial_\nu V_\mu-\frac{1}{2}\eta_{\mu\nu}\partial^\rho V_\rho=0
\eea
where 
\begin{align*}
    V_\rho=\frac{1}{3}\partial^\mu\partial^\nu\partial_\rho h_{\mu\nu}-\Box\partial^\mu h_{\mu\rho}\,.
\end{align*}
$h_{\mu \nu}$ is only defined up to linearized gauge transformations, which can be used to impose the \emph{conformal gauge} $V_\rho=0$ \cite{Riegert:1984hf}. This gauge simplifies \eqref{eq:linearizedBach} to
\be
\Box^2h_{\mu\nu}=0\,,
\ee
which is solved by
\be
h_{\mu\nu}=(A_{\mu\nu}+B_{\mu\nu}n\cdot x)\exp (\im k\cdot x)\,,
\ee
with $k^2=0$ and $n$ chosen to be timelike with $n\cdot k\neq 0$. There is a residual gauge freedom, which shows that $A_{\mu\nu}$ and $B_{\mu\nu}$ contain a total of six degrees of freedom \cite{Adamo:2018srx}. In conformal gauge, they can be distributed so that $A_{\mu\nu}$ contains a pair of helicity $\pm 2$ modes, the \emph{Einstein gravitons}, and a pair of helicity $\pm 1$ modes, sometimes referred to as \emph{photons} \cite{Johansson:2018ues}, while $B_{\mu\nu}$ contains the other pair of helicity $\pm 2$ modes, the \emph{conformal gravitons}. The distribution of the helicity $\pm 1$ modes between $A_{\mu\nu}$ and $B_{\mu\nu}$ is gauge dependent, which makes these modes very subtle objects \cite{Adamo:2018srx}.

The self-dual subsector contains only $3=1+1+1$ of these $6=2+2+2$ degrees of freedom, a structure reminiscent of the three towers found in section \ref{sec:CwAlg}.

\section{Summary of \cw\ and \dw}
Flat-space scattering amplitudes in Einstein gravity form representations of the \cw\ algebra \cite{Strominger:2021mtt, Guevara:2021abz} (see \cite{Raclariu:2021zjz} for a review):
\be
\label{eq:w}
	[w^p_{\bar{m},m},w^q_{\bar{n},n}] = (\bar{m}(q-1)-\bar{n}(p-1))w^{p+q-2}_{\bar{m}+\bar{n},m+n} \,,
\ee
These generators have explicit representations in terms of graviton modes with $\mathfrak{so}(3,1)$ boost weights $(\bar m,m)$ and dilatation weights $2(2-p)$; see equation \eqref{eq:EinsteinCyl} below. They are subject to a wedge condition (see figure \ref{fig:Wedge}):
\be
\label{eq:wedge}
 p\in\{1,\tfrac{3}{2}, 2, \tfrac{5}{2}, \dots \}, \quad \bar{m}\in \{1-p,2-p,\dots, p-2,p-1\}, \quad m\in \mathbb{Z}+p\,.
\ee
\cw\ has a global subalgebra consisting of the four translation generators $P_{\pm \half \pm \half}=w^{3\over 2}_{\pm \half \pm \half}$ and the anti-self-dual Lorentz transformations $L_{\bar m}=w^2_{\bar m,0}$ with $\bar m=0,\pm 1$, sometimes referred to as the chiral Poincare algebra. The self-dual Lorentz transformations are outer automorphisms.

The \cw\ algebra admits a deformation to \dw\
\be
\label{eq:wLambda}
	[w^p_{\bar{m},m},w^q_{\bar{n},n}] = (\bar{m}(q-1)-\bar{n}(p-1))w^{p+q-2}_{\bar{m}+\bar{n},m+n}-\Lambda(m(q-2)-n(p-2))w^{p+q-1}_{\bar{m}+\bar{n},m+n}\,.
\ee
which is a symmetry algebra of gravitational correlators in (A)dS$_4$ with cosmological constant $\Lambda$. The wedge conditions depend on the context \cite{Bittleston:2024rqe,Heuveline:2025nmb,Strominger:2026yrh}; see figure \ref{fig:Wedge} for details. Here, $\bar L_{\bar m}$, $P_{\pm \half \pm \half}$, and $L_m=w^1_{0,m}$ form a global $\mathfrak{so}(3,2)$ or $\mathfrak{so}(4,1)$ subalgebra, the isometry algebra of AdS$_4$ or dS$_4$ depending on the sign of $\Lambda$. In AdS$_4$, one also encounters the alternative wedge condition \cite{Bittleston:2024rqe, Heuveline:2025nmb}:
\be
\label{eq:wedge2}
 p\in \frac{1}{2}\mathbb{Z}, \quad \bar{m}\in \{1-p,2-p,\dots\}, \quad m\in \{p-2,p-1,\dots\}\,. 
\ee

\section{The algebra $\mathcal{C}w_{1+\infty}$}
\label{sec:CwAlg}

In this section, we define the conformal $w$-algebra $\mathcal{C}w_{1+\infty}$ and show that it contains the four-dimensional $\mathfrak{so}(4,2)$ conformal algebra, \cw, and \dw\ as subalgebras.

\subsection{The conformal w-algebra $\mathcal{C}w_{1+\infty}$} 

Here, we construct \Cw, which contains both \cw\ and \dw\ as subalgebras. This is most easily done in projective coordinates. We first define the algebra $\widetilde{\mathcal{C}}w_{1+\infty}$ generated by four towers:
\be
\label{eq:CGen}
C_A[a_1,a_2,a_3,a_4] \,, \qquad A\in\{1,2,3,4\}\,,\qquad a_i\in \mathbb{Z}\,,\qquad\sum_{i=1}^4 a_i=1\,
\ee
These have the commutation relations
\be
\label{eq:confGrav}
[C_A[a_i],C_B[b_i]]=b_AC_B[a_i+b_i-\delta_{Ai}]-a_BC_A[a_i+b_i-\delta_{Bi}]\,.
\ee
The four labels $a_i\in \mathbb{Z}$ can be traded for the three labels ($p,\bar m,m$) 
via a relation like
\be
a_1=p+\bar{m}-1 \qquad a_2=p-\bar{m}-1 \qquad a_3=-p+m+3/2 \qquad a_4=-p-m+3/2\,.
\ee
The wedge conditions analogous to \eqref{eq:wedge} and \eqref{eq:wedge2} read $a_1,a_2\geq 0$ or $a_1,a_3\geq 0$, respectively.

One may verify that this algebra has an ideal generated by
\be
\label{eq:idealH}
H[a_i]=\sum_{A=1}^4 C_A[a_i+\delta_{Ai}]\,,\qquad \sum_{i=1}^4a_i=0\,
\ee
which we denote $\mathfrak{h}$.
The four towers in $\widetilde{\mathcal{C}}w_{1+\infty}$ reduce to three towers after quotienting by $\mathfrak{h}$. 
The conformal $w$-algebra \Cw\ is then defined as the quotient
\be
\label{eq:quotient}
\mathcal{C}w_{1+\infty}=\widetilde{\mathcal{C}}w_{1+\infty}/\mathfrak{h}\,.
\ee
 
We will see in section \ref{sec:TwistorApproach} that these three towers can be interpreted as soft modes of the three positive-helicity particles in the spectrum of conformal gravity: two spin-$+2$ particles, the Einstein graviton and the so-called \emph{conformal graviton}, as well as one spin-$+1$ particle, sometimes referred to as the photon \cite{Adamo:2013tja,Johansson:2017srf,Johansson:2018ues}.

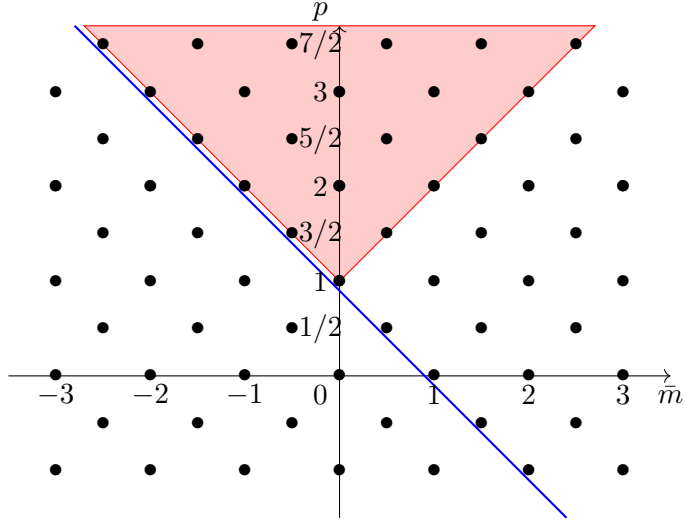
\begin{figure}[t!]
	\begin{center}
\begin{tikzpicture}[scale=1.25]

\filldraw[draw=red, fill=red!20] (0,1) -- (2.7,3.7) -- (-2.7,3.7)  -- cycle;
\draw[->] (-3.5,0) -- (3.5,0);
\node at (3.5,-0.2){$\bar{m}$};
\draw[->] (0,-1.5) -- (0,3.7);

\draw[blue, thick] (-2.8,3.7) -- (2.4,-1.5);

\node at (-0.2,3.85){$p$};
\node at (0,1){$\bullet$};
\node at (-0.2,1){$1$};
\node at (-0.2,1.5){$3/2$};
\node at (0,2){$\bullet$};
\node at (-0.2,2){$2$};
\node at (-0.2,2.5){$5/2$};
\node at (0,3){$\bullet$};
\node at (-0.2,3){$3$};
\node at (-0.2,0.5){$1/2$};
\node at (-0.2,3.5){$7/2$};

\node at (-0.2,-0.2){$0$};
\node at (1,-0.2){$1$};
\node at (2,-0.2){$2$};
\node at (3,-0.2){$3$};
\node at (-1,-0.2){$-1$};
\node at (-2,-0.2){$-2$};
\node at (-3,-0.2){$-3$};

\node at (-3,0){$\bullet$};
\node at (-2,0){$\bullet$};
\node at (-1,0){$\bullet$};
\node at (0,0){$\bullet$};
\node at (3,0){$\bullet$};
\node at (2,0){$\bullet$};
\node at (1,0){$\bullet$};

\node at (-3,1){$\bullet$};
\node at (-2,1){$\bullet$};
\node at (-1,1){$\bullet$};
\node at (0,1){$\bullet$};
\node at (3,1){$\bullet$};
\node at (2,1){$\bullet$};
\node at (1,1){$\bullet$};

\node at (-3,2){$\bullet$};
\node at (-2,2){$\bullet$};
\node at (-1,2){$\bullet$};
\node at (0,2){$\bullet$};
\node at (3,2){$\bullet$};
\node at (2,2){$\bullet$};
\node at (1,2){$\bullet$};

\node at (-3,2){$\bullet$};
\node at (-2,2){$\bullet$};
\node at (-1,2){$\bullet$};
\node at (0,2){$\bullet$};
\node at (3,2){$\bullet$};
\node at (2,2){$\bullet$};
\node at (1,2){$\bullet$};

\node at (-3,3){$\bullet$};
\node at (-2,3){$\bullet$};
\node at (-1,3){$\bullet$};
\node at (0,3){$\bullet$};
\node at (3,3){$\bullet$};
\node at (2,3){$\bullet$};
\node at (1,3){$\bullet$};

\node at (-3,-1){$\bullet$};
\node at (-2,-1){$\bullet$};
\node at (-1,-1){$\bullet$};
\node at (0,-1){$\bullet$};
\node at (3,-1){$\bullet$};
\node at (2,-1){$\bullet$};
\node at (1,-1){$\bullet$};

\node at (2.5,1.5){$\bullet$};
\node at (1.5,1.5){${\bullet}$};
\node at (0.5,1.5){${\bullet}$};
\node at (-0.5,1.5){${\bullet}$};
\node at (-1.5,1.5){${\bullet}$};
\node at (-2.5,1.5){${\bullet}$};

\node at (2.5,2.5){${\bullet}$};
\node at (1.5,2.5){${\bullet}$};
\node at (0.5,2.5){${\bullet}$};
\node at (-0.5,2.5){${\bullet}$};
\node at (-1.5,2.5){${\bullet}$};
\node at (-2.5,2.5){${\bullet}$};

\node at (2.5,3.5){${\bullet}$};
\node at (1.5,3.5){${\bullet}$};
\node at (0.5,3.5){${\bullet}$};
\node at (-0.5,3.5){${\bullet}$};
\node at (-1.5,3.5){${\bullet}$};
\node at (-2.5,3.5){${\bullet}$};

\node at (2.5,0.5){${\bullet}$};
\node at (1.5,0.5){${\bullet}$};
\node at (0.5,0.5){${\bullet}$};
\node at (-0.5,0.5){${\bullet}$};
\node at (-1.5,0.5){${\bullet}$};
\node at (-2.5,0.5){${\bullet}$};

\node at (2.5,-0.5){${\bullet}$};
\node at (1.5,-0.5){${\bullet}$};
\node at (0.5,-0.5){${\bullet}$};
\node at (-0.5,-0.5){${\bullet}$};
\node at (-1.5,-0.5){${\bullet}$};
\node at (-2.5,-0.5){${\bullet}$};

\end{tikzpicture}
\end{center}
	\caption{\emph{The wedge condition \eqref{eq:wedge} means that, for a fixed value of $m$, the generators $w^{p}_{\bar{m},m},S^{a,p}_{\bar{m},m}$ are all located in the red triangle. In the four-index $a_i$ bases of \eqref{eq:CGen}, \eqref{eq:MAB}, this picture is rotated by forty-five degrees, and the wedge condition reads $a_1,a_2\geq0$. The alternative wedge condition \eqref{eq:wedge2} includes only certain generators to the right of the blue line, but it cannot be visualized in two dimensions because the condition depends on both $\bar{m}$ and $m$. In the  $a_i$  basis, \eqref{eq:wedge2} reads $a_1,a_3\geq0$.}} \label{fig:Wedge}
\end{figure}

\subsection{The global subalgebra of \Cw}
\label{subsec:GlobalSubalg}
\label{sec:Subalgebras}

$\mathcal{C}w_{1+\infty}$ contains the global subalgebra $\mathfrak{so}(4,2)$, which is the conformal algebra in four dimensions. It has $15=16-1$ generators. These are the elements $C_A[a_i]$ with $a_1,a_2,a_3,a_4\geq0$. Define a basis by
\be
 T^A_B=C_B[\delta_i^A]\,,
\ee
which indeed consists of fifteen generators because in the quotient algebra \eqref{eq:quotient}, we have $\sum_{A}T^A_A=0$. Equation \eqref{eq:confGrav} implies that 
\be
\label{eq:GlobalComm}
\big[T^A_B,T^C_D\big]
=
\delta^C_B T^A_D
-
\delta^A_D T^C_B\,,
\ee
which are the defining commutation relations of $\mathfrak{sl}(4,\mathbb{C})\cong \mathfrak{so}(6,\mathbb{C})$, from which $\mathfrak{su}(2,2)\cong \mathfrak{so}(4,2)$ is obtained by imposing appropriate Lorentzian reality conditions.  
The familiar generators of boosts, translations, special conformal transformations, and dilations are related to $T^A_B$ by
\bea
\label{eq:Isomo}
\bar L_1&=T^1_2,\qquad
\bar L_0=\frac12\bigl(T^2_2-T^1_1\bigr),\qquad
\bar L_{-1}=-T^2_1,
\\
L_1&=T^3_4,\qquad
L_0=\frac12\bigl(T^4_4-T^3_3\bigr),\qquad
L_{-1}=-T^4_3,
\\
P_{\frac12,\frac12}&=\frac12T^3_2,\qquad
P_{\frac12,-\frac12}=\frac12T^4_2,
\\
P_{-\frac12,\frac12}&=-\frac12T^3_1,\qquad
P_{-\frac12,-\frac12}=-\frac12T^4_1,
\\
K_{\frac12,\frac12}&=2T^1_4,\qquad
K_{\frac12,-\frac12}=-2T^1_3,
\\
K_{-\frac12,\frac12}&=2T^2_4,\qquad
K_{-\frac12,-\frac12}=-2T^2_3,
\\
D&=\frac12\bigl(T^1_1+T^2_2-T^3_3-T^4_4\bigr)\,.
\eea
These have the familiar commutation relations
\bea
\label{eq:so(4,2)}
\left[ L_n, L_m \right] &= (n-m)L_{n+m}
&
\left[ \bar L_{\bar n}, \bar L_{\bar m} \right]
&= (\bar n-\bar m)\bar L_{\bar n+\bar m}
\\[0.8em]
\left[ L_n, P_{\bar r,r} \right]
&= \frac12 (n-2r) P_{\bar r,r+n}
&
\left[ \bar L_{\bar n}, P_{\bar r,r} \right]
&= \frac12 (\bar n-2\bar r) P_{\bar r+\bar n,r}
&
\left[ D, P_{\bar r,r} \right]
&= -P_{\bar r,r}
\\[0.8em]
\left[ L_n, K_{\bar r,r} \right]
&= \frac12 (n-2r) K_{\bar r,r+n}
&
\left[ \bar L_{\bar n}, K_{\bar r,r} \right]
&= \frac12 (\bar n-2\bar r) K_{\bar r+\bar n,r}
&
\left[ D, K_{\bar r,r} \right]
&= K_{\bar r,r}
\eea
\begin{equation*}
\left[ K_{\bar r,r}, P_{\bar s,s} \right]
= -\epsilon_{\bar r,\bar s}\epsilon_{r,s}D
   -\epsilon_{\bar r,\bar s}L_{r+s}
   -\epsilon_{r,s}\bar L_{\bar r+\bar s}
\end{equation*}
where
$\epsilon_{-1/2,1/2}
=
-\epsilon_{1/2,-1/2}
=
1$.
The $7$ generators $\bar{L}_{\bar n},P_{r,\bar{r}}$ are contained in $\mathcal{L}w_{1+\infty}$ as $w^2_{\bar{n},0},w^{3/2}_{\bar{r},r}$ and form a global subalgebra thereof, sometimes referred to as the \emph{chiral Poincare algebra}. Similarly, $\mathcal{L}_\Lambda w_{1+\infty}$ contains the $10$ generators $L_n,\bar{L}_{\bar n},P_{r,\bar{r}}+\Lambda K_{r,\bar{r}}$, which form $\mathfrak{so}(3,2)$ or $\mathfrak{so}(4,1)$ for $\Lambda<0$ and $\Lambda>0$, respectively \cite{Bittleston:2024rqe,Taylor:2023ajd}.

\subsection{AdS$_4$, dS$_4$, and Minkowski space}

Let us now discuss the inclusion $\mathcal{L}_\Lambda w_{1+\infty}\subset \mathcal{C} w_{1+\infty}$. It can be seen most explicitly by introducing
\be
\label{eq:MAB}
M_{AB}[a_i]
=
a_A C_B[a_i-\delta_{iA}]
-
a_B C_A[a_i-\delta_{iB}]\,,
\ee
for $a_1+a_2+a_3+a_4=2$. These generators will also be discussed in more detail in appendix \ref{app:Details} and section \ref{sec:TwistorApproach}. They can be used to define the two spin-$2$ towers which are a subset of the generators \eqref{eq:MAB}
\bea
\label{eq:DefineUV}
u^{p}_{\bar{m},m}
&=\frac{1}{2}\,M_{12}[p+\bar{m}-1,p-\bar{m}-1,-p+m+2,-p-m+2]\\
v^{p}_{\bar{m},m}&=\frac{1}{2}\,M_{34}[p+\bar{m}-1,p-\bar{m}-1,-p+m+2,-p-m+2]\,.
\eea
They obey the following commutation relations among themselves:
\bea
\label{eq:uvAlg}
[u^{p}_{\bar{m},m},u^{q}_{\bar{n},n}]&=(\bar{m}(q-1)-\bar{n}(p-1))u^{p+q-2}_{\bar{m}+\bar{n},m+n}\\
[v^{p}_{\bar{m},m},v^{q}_{\bar{n},n}]&=-(m(q-2)-n(p-2))v^{p+q-1}_{\bar{m}+\bar{n},m+n}\,,
\eea
and further
\be
[u^p_{\bar m,m},v^q_{\bar n,n}]
+
[v^p_{\bar m,m},u^q_{\bar n,n}]
=
\bigl(\bar m(q-1)-\bar n(p-1)\bigr)
v^{p+q-2}_{\bar m+\bar n,m+n}
-
\bigl(m(q-2)-n(p-2)\bigr)
u^{p+q-1}_{\bar m+\bar n,m+n}\,.
\ee
Using this, it is easy to see that for arbitrary $\Lambda$, the generators
\be w^p_{\bar{m},m}=u^{p}_{\bar{m},m}+\Lambda v^{p}_{\bar{m},m}
\ee
form a closed subalgebra of \Cw\ that obeys the defining relation of $\mathcal{L}_\Lambda w_{1+\infty}$:
\bea
\label{eq:AdSAlg}
&[w^p_{\bar{m},m},w^q_{\bar{n},n}]
&=(\bar{m}(q-1)-\bar{n}(p-1))w^{p+q-2}_{\bar{m}+\bar{n},m+n}-\Lambda(m(q-2)-n(p-2))w^{p+q-1}_{\bar{m}+\bar{n},m+n}\,.
\eea
A special case of this construction is $\Lambda=0$, which also shows that $\mathcal{L}w_{1+\infty}\subset \mathcal{C}w_{1+\infty}$.

As discussed above, $\mathcal{L}_\Lambda w_{1+\infty}$ intersects $\mathfrak{so}(4,2)$ in $\mathfrak{so}(3,2)$, $\mathfrak{so}(4,1)$, or the chiral Poincare algebra for $\Lambda<0$, $\Lambda>0$, and $\Lambda=0$, respectively.

$\mathcal{C}w_{1+\infty}$ admits the automorphism
\be
\label{eq:ConformalInversion0}
i(C_A[a_1,a_2,a_3,a_4])=C_{A+2}[a_3,a_4,a_1,a_2]\,,
\ee
where addition in the $A$ label is taken modulo $4$. This automorphism corresponds to conformal inversion.
For $\Lambda\neq 0$, equation \eqref{eq:ConformalInversion0} acts on $w^p_{\bar{m},m}$ by
\be
i\big(u^{p}_{\bar{m},m}+\Lambda v^{p}_{\bar{m},m}\big)= \Lambda\big( u^{3-p}_{m,\bar{m}}+\Lambda^{-1} v^{3-p}_{m,\bar{m}}\big)\,.
\ee
Note that this conformal inversion violates the standard wedge condition \eqref{eq:wedge}, whereas it preserves the alternative wedge condition \eqref{eq:wedge2}.\footnote{In the case of Euclidean AdS$_4$, these two wedge conditions correspond to the ball model and the half-plane model of Euclidean AdS$_4$, respectively \cite{Bittleston:2024rqe,Heuveline:2025nmb,Strominger:2026yrh}. The conformal inversion $x^{\alpha\dal}\mapsto x^{\alpha\dal}/x^2$ violates the defining condition of the ball model, $x^2\leq 1$, but preserves the half-plane condition, $n\cdot x\geq 0$.}

\subsection{The Einstein cylinder}

The maximally symmetric spacetimes we have considered so far are all conformally flat and Einstein. There are other physically interesting conformally flat spacetimes that are not Einstein. Examples include FLRW metrics, AdS$_2\times S^2$, and the $S^3\times\mathbb{R}$ Einstein cylinder EC$_4$, which we briefly discuss in this section.

The metric of EC$_4$ is
\be ds^2=-\dif t^2+\dif \Omega_3^2\,,\ee
which solves the Einstein field equations with a traceless diagonal stress tensor $T^{\text{EC}}_{\mu\nu}$. This can be sourced by a perfect fluid or by pressureless dust plus a cosmological constant.

We still expect that, because EC$_4$ is conformally flat, there should be an infinite-dimensional subalgebra of $\mathcal{C}w_{1+\infty}$ that describes Einstein gravity in the presence of $T^{\text{EC}}_{\mu\nu}$ on EC$_4$. Because the Einstein cylinder is still highly symmetric, it is possible to find a subalgebra of $\mathcal{C}w_{1+\infty}$ that contains the isometries of EC$_4$ as a global subalgebra.
These are generated by the $7$-dimensional Lie algebra
\be
\mathfrak{so}(4)\oplus \mathfrak{u}(1)\,,
\ee
where $\mathfrak{so}(4)$ acts on the $S^3$ factor and $\mathfrak{u}(1)$ acts on the $\mathbb{R}$ factor. The algebra $\mathfrak{so}(4)\times \mathfrak{u}(1)$ can be embedded into $\mathfrak{so}(4,2)$, where it is realized by the seven generators
\be
\label{eq:su2u1}
\bar{L}_{\bar{n}}=u^2_{\bar{n},0}\,,\quad L_{n}= v^1_{0,n}\,,\quad D \,,
\ee
where $n, \bar{n}\in \{-1,0,1\}$.
We want to complete \eqref{eq:su2u1} into an infinite-dimensional algebra that contains only one tower of generators; in particular, for generic weights, it should not contain multiple generators with the same dilatation weight and the same $L_0$ and $\bar{L}_{0}$ weights. We have found that a natural such algebra is given by the generators\footnote{The isometries of the Einstein cylinder are not necessarily inner automorphisms and they could also act as outer automorphisms, meaning that the first line of \eqref{eq:EC} would be removed from the generators. This happens in flat space where only the chiral Poincare algebra acts by inner automorphisms \cite{Strominger:2021mtt}.}
\bea
\label{eq:EC}
&v^1_{0,1}\,,\quad v^1_{0,0}\,,\quad v^1_{0,-1}\,,\quad D \,,\\
&u^p_{\bar{m},m}\qquad \text{for any $p\geq2$}\,.
\eea
The commutators of these generators can then be read off from \eqref{eq:uvAlg} and \eqref{eq:confGrav}. They are given by
\bea
\label{eq:EinsteinCyl}
[u^{p}_{\bar{m},m},u^{q}_{\bar{n},n}]&=(\bar{m}(q-1)-\bar{n}(p-1))\,u^{p+q-2}_{\bar{m}+\bar{n},m+n}\\
[D,u^p_{\bar{m},m}]&=2(p-2)\,u^p_{\bar{m},m}\\
[v^1_{0,n},u^p_{\bar{m},m}]&=-(n(p-2)+m)\,u^p_{\bar{m},m+n}\,.
\eea
We conjecture that this algebra describes Einstein gravity in the presence of $T_{\mu\nu}^{\text{EC}}$ on an EC$_4$ background, and we hope to prove this statement in future work.

\section{Coupling to the $S$-algebra}
In Yang-Mills theory, in close analogy to the gravitational case, there is a tower of soft theorems whose commutator algebra is the so-called $S$-algebra:
\be
\label{eq:S}
[S^{p,a}_{\bar{m},m},S^{q,b}_{\bar{n},n}] = -\im f^{ab}_c\,S^{p+q-1,c}_{\bar{m}+\bar{n},m+n}\,,
\ee
where $a,b,c$ are color indices. The action of the conformal algebra $\mathfrak{so}(4,2)$ on the $S$-algebra was derived in \cite{Sheta:2025oep}, the action of \cw\ was derived in \cite{Strominger:2021mtt}, and the action of $\mathcal{L}_\Lambda w_{1+\infty}$ was derived in \cite{Bittleston:2024rqe}. Here, we generalize and unify these three results by deriving the action of $\mathcal{C}w_{1+\infty}$ on the $S$-algebra.

Let us introduce $S$-algebra generators in a basis labeled by four integers\footnote{The $S$-algebra generators $S^a[b_i]$ with the standard wedge condition are defined for $b_1,b_2\in \mathbb{N}$ and $b_3,b_4\in \mathbb{Z}$ with $b_1+b_2+b_3+b_4=0$.} \cite{Costello:2022wso}, related to the standard celestial basis by
\be
S^{c,p}_{\bar{m},m}=S^c[p+\bar{m}-1,p-\bar{m}-1,-p+m+1,-p-m+1]\,.
\ee
The generators $C_A$ of $\mathcal{C}w_{1+\infty}$ then act simply as
\be
\label{eq:CScoupling}
[C_A[a_i],S^c[b_i]]=b_A S^c[a_i+b_i-\delta_{Ai}]\,.
\ee
This leads to the action of the generators $u^{p}_{\bar{m},m}$ and $v^{p}_{\bar{m},m}$ on the $S$-algebra generators:
\bea
[u^{p}_{\bar{m},m},S^{a,q}_{\bar{n},n}]&=(\bar{m}(q-1)-\bar{n}(p-1))S^{a,p+q-2}_{\bar{m}+\bar{n},m+n}\\
[v^{p}_{\bar{m},m},S^{a,q}_{\bar{n},n}]&=(n(p-2)-m(q-1))S^{a,p+q-1}_{\bar{m}+\bar{n},m+n}\,,
\eea
which is consistent with \cite{Bittleston:2024rqe}.

\section{Generating \Cw\ by the action of $\mathfrak{so}(4,2)$ on $\mathcal{L}w_{1+\infty}$ or $\mathcal{L}_\Lambda w_{1+\infty}$}
\label{sec:so42Action}

In this section, we show that acting with $\mathfrak{so}(4,2)$ on either $\mathcal{L}w_{1+\infty}$ or \dw\ generates the larger algebra $\mathcal{C}w_{1+\infty}$.
Some explicit calculational details are deferred to appendix \ref{app:Details}.

\subsection{Yang-Mills theory}
We first consider the $S$-algebra. Since $\mathfrak{so}(4,2)$ belongs to \Cw, the conformal action on the $S$-algebra is contained in equation \eqref{eq:CScoupling}. Explicitly, it is known \cite{Sheta:2025oep} that $\mathfrak{so}(4,2)$ acts on the S-algebra in our conventions through
\bea
\label{eq:ConfScoupled}
\left[ K_{\bar r,r}, S_{\bar m,m}^{p,a} \right]
&= -4r\big((p-1)+2r m\big)
S_{\bar m+\bar r,\,m+r}^{p+\frac12,a}\\
\left[ P_{\bar r,r}, S_{\bar m,m}^{p,a} \right]
&= \bar{r}\bigl((p-1)-2\bar r\bar m\bigr)
S_{\bar m+\bar r,\,m+r}^{p-\frac12,a}\\
\left[ D, S_{\bar m,m}^{p,a} \right]
&= 2(p-1)S_{\bar m,m}^{p,a}\\
\left[ L_n, S_{\bar m,m}^{p,a} \right]
&= \bigl(n(1-p)-m\bigr)S_{\bar m,\,m+n}^{p,a}\\
\left[ \bar L_{\bar n}, S_{\bar m,m}^{p,a} \right]
&= \bigl(\bar n(p-1)-\bar m\bigr)
S_{\bar m+\bar n,\,m}^{p,a}\,,
\eea
which can also be read off from \eqref{eq:CScoupling}. This closure of the $S$-algebra under the $\mathfrak{so}(4,2)$ action is implied by the conformal invariance of Yang-Mills theory.

\subsection{Gravity}
Like Yang-Mills theory, classical conformal gravity is conformally invariant. Hence its symmetry algebra \Cw\ is closed under $\mathfrak{so}(4,2)$, whose action in the gravity case is an inner rather than an outer automorphism.

Einstein gravity, on the other hand, is not conformally invariant. In particular, we do not expect $\mathcal{L}w_{1+\infty}$ to be invariant under the action of $\mathfrak{so}(4,2)$, as can be readily verified from section \ref{sec:CwAlg}. However, all solutions of Einstein gravity are solutions of conformal gravity, which is conformally invariant. Therefore, the $\mathfrak{so}(4,2)$ action on the generators of \cw\ or \dw, viewed as graviton wavefunctions, produces a conformal-gravity wavefunction: a spin-$2$ Einstein graviton, a spin-$2$ ghost graviton, or a photon. In this section, we show that repeated action gives all of \Cw.

%

\subsection{Acting with $\mathfrak{so}(4,2)$ on $\mathcal{L}w_{1+\infty}$}

For $\Lambda=0$, the chiral Poincare algebra generated by the $7$ generators $\bar{L}_{\bar n},P_{r,\bar{r}}$ is contained in $\mathcal{L}w_{1+\infty}$. In particular, the commutator of any of these elements with a $w^p_{\bar{m},m}$ generates another $w^p_{\bar{m},m}$.
However, other elements of $\mathfrak{so}(4,2)$, such as the special conformal transformations, do not preserve the generators $w^p_{\bar{m},m}$. 

We can formally generate a new Lie algebra that contains both $\mathfrak{so}(4,2)$ and $\mathcal{L}w_{1+\infty}$ by including new generators $[T^A_B,w^p_{\bar{m},m}]$ and constraining their commutation relations with other generators by requiring the Jacobi identity. For instance, further generators $\left[
w^p_{\bar{m},m},\left[w^q_{\bar{n},n},T^A_B\right]\right]$ have to be included and obey the relation
\be
\left[w^q_{\bar{n},n},
 \left[T^A_B,w^p_{\bar{m},m}\right]\right]+\left[
w^p_{\bar{m},m},\left[w^q_{\bar{n},n},T^A_B\right]\right]
=
\left(\bar{n}(p-1)-\bar{m}(q-1)\right)
\left[T^A_B,
 w^{p+q-2}_{\bar{m}+\bar{n},m+n}\right]\,,
\ee
and similarly with another $T^C_D$:
\be
\left[T^C_D,
 \left[T^A_B,w^p_{\bar{m},m}\right]\right]+\left[T^A_B,
 \left[w^p_{\bar{m},m},T^C_D\right]\right]
=
\delta^A_D\left[T^C_B,w^p_{\bar{m},m}\right]
-\delta^C_B\left[T^A_D,w^p_{\bar{m},m}\right]\,.
\ee
Commutators such as
$[[T^A_B,w^p_{\bar{m},m}],[T^C_D,w^q_{\bar{n},n}]]$
obey more complicated relations that require us to include more generators. They satisfy
\bea
\left[\left[T^A_B,w^p_{\bar{m},m}\right],
      \left[T^C_D,w^q_{\bar{n},n}\right]\right]
={}&
\left[T^A_B,
 \left[w^q_{\bar{n},n},
  \left[T^C_D,w^p_{\bar{m},m}\right]\right]\right]
\\
&+\left(\bar{m}(q-1)-\bar{n}(p-1)\right)
 \left[T^A_B,
  \left[T^C_D,
   w^{p+q-2}_{\bar{m}+\bar{n},m+n}\right]\right]
\\
&-\delta^C_B
 \left[w^p_{\bar{m},m},
  \left[T^A_D,w^q_{\bar{n},n}\right]\right]
+\delta^A_D
 \left[w^p_{\bar{m},m},
  \left[T^C_B,w^q_{\bar{n},n}\right]\right]
\\
&-\left[w^p_{\bar{m},m},
 \left[T^C_D,
  \left[T^A_B,w^q_{\bar{n},n}\right]\right]\right].
\eea
and so on. Thus, this procedure formally gives us a new algebra that, by construction, contains both $\mathfrak{so}(4,2)$ and $\mathcal{L}w_{1+\infty}$.
In appendix \ref{app:Details}, we will see that all of the generators $C_A[a_i]$ can be built from such commutators of $T^A_B$ with $w^p_{\bar{m},m}$ and hence that this algebra gives rise to $\mathcal{C}w_{1+\infty}$. For instance, it turns out that the following complicated-looking combination
\be
\label{eq:generatingC1}
-\frac12\left(
(a_3+a_4+3)\,
w^{\frac32+\frac{a_1+a_2}{2}}_{
\frac{a_1-a_2-1}{2},\,
\frac{a_3-a_4}{2}}
+
\left[
w^{1+\frac{a_1+a_2}{2}}_{
\frac{a_1-a_2}{2},\,
\frac{a_3-a_4+1}{2}},
T^2_3
\right]
+
\left[
w^{1+\frac{a_1+a_2}{2}}_{
\frac{a_1-a_2}{2},\,
\frac{a_3-a_4-1}{2}},
T^2_4
\right]
\right)
\ee
can be identified with $C_1[a_i]$, and similar formulas express all the $C_A[a_i]$ in terms of nested commutators of $T^A_B$ with $w^p_{\bar{m},m}$ as discussed in appendix \ref{app:Details}. Note that the two global generators needed in \eqref{eq:generatingC1} are special conformal generators:
\be
T^2_3=-\frac{1}{2}K_{-1/2,-1/2}\qquad 
T^2_4=\frac{1}{2}K_{-1/2,1/2}\,.
\ee

Strictly speaking, the Jacobi identity alone does not uniquely determine
an abstract Lie algebra once the mixed commutators
$[T^A_B,w^p_{\bar m,m}]$ are introduced: it imposes consistency
conditions on their further commutators, but may still allow additional central extensions or quotients. In the following, by the
algebra generated by the action of $\mathfrak{so}(4,2)$ on
$\mathcal{L}w_{1+\infty}$ we mean the smallest Lie algebra in the
concrete realization considered in section \ref{sec:TwistorApproach}. In this realization, all successive commutators
automatically satisfy the Jacobi identity, and
appendix~\ref{app:Details} shows that they generate every
$C_A[a_i]$. In this concrete sense, the Lie closure obtained by the
above procedure is precisely $\mathcal{C}w_{1+\infty}$.

\subsection{Acting with $\mathfrak{so}(4,2)$ on $\mathcal{L}_\Lambda w_{1+\infty}$}
\label{subsec:LLambda}

Acting with $\mathfrak{so}(4,2)$ on $\mathcal{L}_\Lambda w_{1+\infty}$ also leads to \Cw. Recall that $\mathcal{L}_\Lambda w_{1+\infty}$ is defined by generators $w^p_{\bar{m},m}$ obeying
\bea
\label{eq:AdSAlg1}
&[w^p_{\bar{m},m},w^q_{\bar{n},n}]
&=(\bar{m}(q-1)-\bar{n}(p-1))w^{p+q-2}_{\bar{m}+\bar{n},m+n}-\Lambda(m(q-2)-n(p-2))w^{p+q-1}_{\bar{m}+\bar{n},m+n}\,.
\eea
As in the case $\Lambda=0$, $\mathcal{L}_\Lambda w_{1+\infty}$ is not conformally invariant, and the commutator $[T^A_B,w^p_{\bar{m},m}]$ takes the form of another $w^p_{\bar{m},m}$ only if $T^A_B$ is contained in the appropriate $\mathfrak{so}(3,2)$ or $\mathfrak{so}(4,1)$ subalgebra.
Once again, we need to formally include new generators $[T^A_B,w^p_{\bar{m},m}]$, which now obey the following deformed relations:
\bea
\left[w^q_{\bar{n},n},
 \left[T^A_B,w^p_{\bar{m},m}\right]\right]
={}&
\left[\left[w^q_{\bar{n},n},T^A_B\right],
 w^p_{\bar{m},m}\right]
+\left(\bar{n}(p-1)-\bar{m}(q-1)\right)
 \left[T^A_B,
  w^{p+q-2}_{\bar{m}+\bar{n},m+n}\right]
\\
&+\Lambda\left(m(q-2)-n(p-2)\right)
 \left[T^A_B,
  w^{p+q-1}_{\bar{m}+\bar{n},m+n}\right],
\\
\left[T^C_D,
 \left[T^A_B,w^p_{\bar{m},m}\right]\right]
={}&
\delta^A_D\left[T^C_B,w^p_{\bar{m},m}\right]
-\delta^C_B\left[T^A_D,w^p_{\bar{m},m}\right]
+\left[T^A_B,
 \left[T^C_D,w^p_{\bar{m},m}\right]\right]\,,
\eea

and
\bea
\left[\left[T^A_B,w^p_{\bar{m},m}\right],
      \left[T^C_D,w^q_{\bar{n},n}\right]\right]
={}&
\left[T^A_B,
 \left[w^q_{\bar{n},n},
  \left[T^C_D,w^p_{\bar{m},m}\right]\right]\right]
\\
&+\left(\bar{m}(q-1)-\bar{n}(p-1)\right)
 \left[T^A_B,
  \left[T^C_D,
   w^{p+q-2}_{\bar{m}+\bar{n},m+n}\right]\right]
\\
&-\Lambda\left(m(q-2)-n(p-2)\right)
 \left[T^A_B,
  \left[T^C_D,
   w^{p+q-1}_{\bar{m}+\bar{n},m+n}\right]\right]
\\
&-\delta^C_B
 \left[w^p_{\bar{m},m},
  \left[T^A_D,w^q_{\bar{n},n}\right]\right]
+\delta^A_D
 \left[w^p_{\bar{m},m},
  \left[T^C_B,w^q_{\bar{n},n}\right]\right]
\\
&-\left[w^p_{\bar{m},m},
 \left[T^C_D,
  \left[T^A_B,w^q_{\bar{n},n}\right]\right]\right]\,.
\eea
With these relations at hand, it is straightforward to see that the two spin-$2$ towers of section \ref{sec:CwAlg} can be recovered. Indeed, using the dilatation $D=\frac{1}{2}(T^1_1+T^2_2-T^3_3-T^4_4)$ (as discussed in \eqref{eq:Isomo}), the two towers
\bea
\label{eq:RecoverUV}
u^p_{\bar m,m}
&=(p-1)w^p_{\bar m,m}
-\frac12[D,w^p_{\bar m,m}]\,,\\
v^p_{\bar m,m}
&=\frac{1}{2\Lambda}
\bigl([D,w^p_{\bar m,m}]
-2(p-2)w^p_{\bar m,m}\bigr)
\eea
obey the correct relation \eqref{eq:uvAlg}:
\bea
[u^{p}_{\bar{m},m},u^{q}_{\bar{n},n}]&=(\bar{m}(q-1)-\bar{n}(p-1))u^{p+q-2}_{\bar{m}+\bar{n},m+n}\\
[v^{p}_{\bar{m},m},v^{q}_{\bar{n},n}]&=-(m(q-2)-n(p-2))v^{p+q-1}_{\bar{m}+\bar{n},m+n}\,.
\eea
Thus, for $\Lambda\neq0$, a single action with the dilatation generator $D$ already generates
the $u$-tower and $v$-tower separately. In particular, the generated
algebra contains the complete $u$-tower, which is the undeformed
algebra $\mathcal{L}w_{1+\infty}$. Taking commutators of elements of the $u$ and $v$ towers and acting with the remaining generators
of $\mathfrak{so}(4,2)$ then generates all of
$\mathcal{C}w_{1+\infty}$ by the argument of appendix \ref{app:Details}. Since both
$\mathfrak{so}(4,2)$ and $\mathcal{L}_\Lambda w_{1+\infty}$ are
themselves contained in $\mathcal{C}w_{1+\infty}$, their conformal
closure cannot generate anything larger. We therefore obtain the same
algebra $\mathcal{C}w_{1+\infty}$ from this procedure, independently of the value of
$\Lambda$.

\section{Twistor analysis}
\label{sec:TwistorApproach}
\label{sec:SDCG}

In this section, we derive $\mathcal{C}w_{1+\infty}$ from a complementary perspective, namely as the algebra of gauge symmetries of a certain holomorphic theory on twistor space. In particular, we focus on \emph{self-dual conformal gravity} as defined in section \ref{sec:confGrav} and its uplift to a local action on twistor space, so-called \emph{BN theory}. This self-dual subsector is a classically integrable field theory with an infinite-dimensional \Cw\ symmetry algebra that is manifest in BN theory. 

\subsection{The twistor action and its symmetries}
 
It is known that \eqref{eq:SDCG} is classically equivalent to a local action on twistor space \cite{Mason:2005zm}. Here, we review this action, which is based heavily on the \emph{non-linear graviton construction} \cite{Penrose:1976js}, and show that it manifestly contains an infinite-dimensional gauge symmetry given precisely by $\mathcal{C}w_{1+\infty}$, as defined in section \ref{sec:CwAlg}. The non-linear graviton construction is a precise mathematical equivalence between certain complex threefolds, known as \emph{curved twistor spaces}, and conformal classes of four-dimensional self-dual spacetimes.

\begin{theorem}[Non-linear graviton correspondence]
\label{thm:nonlin}
There is a one-to-one correspondence between:

\begin{itemize}
    \item[(i)] Complex three-manifolds $\mathcal{PT}$ containing a four-parameter family of holomorphically embedded rational curves 
    $X \cong \mathbb{CP}^1$ with normal bundle
    \begin{equation}
    N_{X/\mathbb{PT}} \cong \mathcal{O}(1) \oplus \mathcal{O}(1)\,,
    \end{equation}

    \item[(ii)] Complex four-dimensional manifolds $M$ equipped with a self-dual conformal structure $[g]$,
    \begin{equation}
    \Psi_{\alpha\beta \gamma \delta} = 0\,.
    \end{equation}
\end{itemize}
\end{theorem}

Through this correspondence, points $x\in M$ correspond to rational curves $X\subset\mathcal{PT}$, twistors $Z\in\mathcal{PT}$ correspond to \emph{$\alpha$-surfaces} in $M$, and the conformal structure on $M$ is uniquely encoded in the complex structure of $\mathcal{PT}$ (see \cite{Adamo:2017qyl} for a review of general twistor theory).

Recasting this theorem in a language suitable for quantum field theory involves perturbing around a background that we take to be (the conformal class of) complexified Minkowski space $(\mathbb{M}_{\mathbb{C}},\eta_{\mu\nu})$. Perturbing this conformal structure of Minkowski space, $[\eta_{\mu\nu}]\mapsto[\eta_{\mu\nu}+\varepsilon h_{\mu\nu}]$, corresponds to perturbing the complex structure on $\mathbb{PT}$. Concretely, the latter means deforming what we mean by holomorphic objects:
\be
\bar{\partial}\rightarrow\bar{\nabla}=\bar{\partial}+\beta\,,
\ee
where $\beta\in\Omega^{0,1}(\mathbb{PT},T^{1,0}\mathbb{PT})$ is a so-called \emph{Beltrami differential}, which will be the dynamical field on twistor space.

Not every deformation of $[\eta_{\mu\nu}]$ is self-dual and, correspondingly, not every $\beta$ leads to a complex manifold. The condition that $\bar{\nabla}=\bar{\partial}+\beta$ defines an integrable complex structure is equivalent to the vanishing of the so-called \emph{Nijenhuis tensor}:
\be
\label{eq:MaurerCartan}
0=N=(\bar{\partial}+\beta)^2=\bar{\partial}\beta+\frac{1}{2}[\beta,\beta]\in  \Omega^{0,2}(\mathbb{PT}, T^{1,0}\mathbb{PT})\,.
\ee

A natural twistor action, often referred to as \emph{BN theory}, arises by imposing equation \eqref{eq:MaurerCartan} with a Lagrange multiplier $B\in\Omega^{0,1}(\mathbb{PT},\mathcal{O}(-4)\otimes\Omega^1)$:
\be
\label{eq:BNaction}
\int_{\mathbb{PT}} \text{D}^3Z \,B\wedge N\,.
\ee
The action \eqref{eq:BNaction} is defined using the weighted holomorphic $(3,0)$-form:
\be
    \Dif^3Z = \frac{1}{4!}\epsilon_{abcd}Z^a\dif Z^b\wedge\dif Z^c\wedge\dif Z^d \in \Omega^{3,0}(\mathbb{PT}, \mathcal{O}(4))\,.
\ee
Note that the action \eqref{eq:BNaction} is invariant under the gauge transformations:
\bea
\label{eq:GaugeSymmBN1}
\delta_\chi \beta &= \bar{\partial}\chi + [\beta,\chi], \\
\delta_\chi B &= [B,\chi]\,,
\eea
where $\chi\in\Omega^{0,0}(\mathbb{PT},T^{1,0}\mathbb{PT})$. There is also an additional transformation familiar from BF theories:
\bea
\label{eq:GaugeSymmBN2}
\delta_\xi B &= \bar{\partial}\xi + [\beta,\xi], \\
\delta_\xi \beta &= 0\,,
\eea
where $\xi\in\Omega^{0,0}(\mathbb{PT},\mathcal{O}(-4)\otimes\Omega^1)$.
Using these symmetries, it was proven in \cite{Mason:2005zm} that the twistor action \eqref{eq:BNaction} is classically equivalent to the spacetime action \eqref{eq:SDCG} \cite{Adamo:2013tja}.
    
The equivalence between these two actions can be viewed, in particular, as an off-shell version of the non-linear graviton construction, since restricting to solutions of the corresponding equations of motion recovers the
one-to-one correspondence of theorem \ref{thm:nonlin}. The advantage of the action \eqref{eq:BNaction} is that it has a manifest infinite-dimensional gauge symmetry at the classical level; it is precisely this symmetry that underlies the classical integrability of \eqref{eq:SDCG}. The algebra formed by these gauge transformations is precisely $\mathcal{C}w_{1+\infty}$, discussed in section \ref{sec:CwAlg}. Let us show this explicitly by considering explicit modes for the vector fields $\chi$\footnote{The vector field $\xi$ leads to generators corresponding to negative helicity states \cite{Costello:2022wso}. The corresponding extension of the algebra can be derived in full analogy to the discussion below but will not be further discussed here.}.

\subsection{Twistor-coordinate realization of $\mathcal{C}w_{1+\infty}$}
\label{subsec:Twistors}

We wish to consider a basis of modes for $\chi\in\Omega^{0,0}(\mathbb{PT},T^{1,0}\mathbb{PT})$ and compute their algebra explicitly. Consider the homogeneous twistor coordinates $Z^A\in\mathbb{PT}\subset \mathbb{CP}^3$ with $A\in\{1,2,3,4\}$\footnote{Below, we will often use the conventional decomposition
$Z^A=(\mu^\da,\lambda_\al)\,.$}.
In analogy to \cite{Costello:2022wso, Adamo:2021lrv}, it is natural to consider the following generators for $\chi$:
\be
\label{eq:TwistorGaugeModes}
C_A[a_1,a_2,a_3,a_4]= (Z^1)^{a_1}(Z^2)^{a_2}(Z^3)^{a_3}(Z^4)^{a_4}\frac{\partial}{\partial Z^A}\,,
\ee
where $A\in\{1,2,3,4\}$, $a_i\in\mathbb{Z}$, and  $\sum_{i=1}^4a_i=1$ is required for \eqref{eq:TwistorGaugeModes} to be weightless. Moreover, an appropriate wedge condition (such as $a_1,a_2\geq0$ or $a_1,a_3\geq0$) is required in order for the vector fields to have an appropriate pole structure as discussed in \cite{Adamo:2021lrv, Costello:2022wso}.
\eqref{eq:TwistorGaugeModes} can be checked immediately to obey the commutation relations \eqref{eq:confGrav}. Moreover, the ideal $\mathfrak{h}$ of equation \eqref{eq:idealH} contains all vector fields that are proportional to the total homogeneity operator. 
Any multiple of the homogeneity operator is vertical and projects to zero in $T\mathbb{PT}$ so that it is natural to identify any two vector fields which differ by an element in $\mathfrak{h}$.
This means that the abstract generators \eqref{eq:CGen} can indeed be identified with the modes of twistorial gauge transformations \eqref{eq:TwistorGaugeModes}.

\paragraph{Einstein gravity from conformal gravity}
The one-to-one correspondence in theorem \ref{thm:nonlin} only makes a statement about the conformal class of the spacetime metric. In order to obtain a self-dual Einstein metric in the conformal class of self-dual metrics, certain $\Lambda$-dependent extra data are required on twistor space, the so-called infinity twistor, which is discussed in more detail in the literature \cite{Penrose:1985bww, Penrose:1986ca, Adamo:2017qyl, Adamo:2021lrv, Heuveline:2025nmb}. For $\Lambda=0$, in standard conventions \cite{Adamo:2017qyl} this infinity twistor amounts to the following Poisson bracket 
\be
\{\quad, \quad\}_{12}=I^{AB}_{[12]}\frac{\partial}{\partial Z^A}\wedge \frac{\partial}{\partial Z^B}\,,\qquad I^{AB}_{[12]}=\delta^A_{1}\delta^B_{2}-\delta^A_{2}\delta^B_{1}\,.
\ee
It is then natural to represent the spin-$2$ Einstein graviton tower $u^p_{\bar{m},m}$ by the following Hamiltonian vector fields \cite{Adamo:2021lrv}:
\be
\label{eq:wHam}
u^p_{\bar{m},m}=\bigg\{\frac{(\mu^{\dot 0})^{p+\bar{m}-1}(\mu^{\dot 1})^{p-\bar{m}-1}}{2\lambda_0^{p-m-2}\lambda_1^{p+m-2}},\qquad \bigg\}_{12}\,,
\ee
which can be easily checked to obey the undeformed algebra \eqref{eq:w}.
In this language, it becomes intuitive why $\mathfrak{so}(4,2)$ does not respect the form of a generator $u^p_{\bar{m},m}$. A generator $T^A_B$ acts on the Hamiltonian as well as on the Poisson bracket. The matrix $I_{[12]}$ singles out $Z^1$ and $Z^2$ and hence breaks $\mathfrak{su}(2,2)\cong\mathfrak{so}(4,2)$. The algebra $\mathfrak{so}(4,2)$ transforms $I_{[12]}$ into $I_{[CD]}$ for any values of $C,D \in \{1,2,3,4\}$, where
\be
I^{AB}_{[CD]}=\delta^A_{C}\delta^B_{D}-\delta^A_{D}\delta^B_{C}\,,
\ee
which defines the further Poisson brackets:
\be
\{\quad, \quad\}_{CD}=I^{AB}_{[CD]}\frac{\partial}{\partial Z^A}\wedge \frac{\partial}{\partial Z^B}\,.
\ee
We will show in appendix \ref{app:Details} that the space of all Hamiltonian vector fields for any of the Poisson brackets $\{\quad,\quad\}_{CD}$ is a generating system\footnote{It does not form a basis, since the relations \eqref{eq:MABRelation} hold between Hamiltonian vector fields for different Hamiltonians.} for the algebra \Cw .

The $\Lambda$-deformed algebra can again be represented as an algebra of Hamiltonian vector fields, where the Poisson bracket is now deformed to\footnote{Formally speaking, this is a (weighted, holomorphic) Jacobi bracket and not a Poisson bracket, as discussed in \cite{Heuveline:2025nmb,Bittleston:2024rqe}.
}
\be
\label{eq:wLambdaGen}
w^p_{\bar{m},m}=\bigg\{\frac{(\mu^{\dot 0})^{p+\bar{m}-1}(\mu^{\dot 1})^{p-\bar{m}-1}}{2\lambda_0^{p-m-2}\lambda_1^{p+m-2}},\qquad \bigg\}_{12}+\Lambda \bigg\{\frac{(\mu^{\dot 0})^{p+\bar{m}-1}(\mu^{\dot 1})^{p-\bar{m}-1}}{2\lambda_0^{p-m-2}\lambda_1^{p+m-2}},\qquad \bigg\}_{34}\,.
\ee
Equation \eqref{eq:wLambdaGen} can be checked directly to satisfy the algebra \eqref{eq:AdSAlg1}. As in the case $\Lambda=0$, this Poisson bracket breaks conformal invariance. More specifically,
\be
\{\quad, \quad\}_{12}+\Lambda \{\quad, \quad\}_{34}
\ee
breaks $\mathfrak{su}(2,2)\cong\mathfrak{so}(4,2)$ to $\mathfrak{so}(3,2)$ or $\mathfrak{so}(4,1)$\footnote{These are both real forms of $\mathfrak{sp}(4,\mathbb{C})\cong\mathfrak{so}(5,\mathbb{C})$, which by definition preserves the standard symplectic structure in four dimensions. $I_{12}+\Lambda I_{34}$ is equivalent to this standard symplectic structure over $\mathbb{C}$. Over $\mathbb{R}$, different signs of $\Lambda$ yield inequivalent real Lie algebras.} depending on the sign of $\Lambda$.

Exchanging the two-component spinors $\lambda_\alpha$ and $\mu^{\dot{\alpha}}$ that comprise the four-vector $Z^A$ is known to be related to conformal inversion through the twistor correspondence. This can be derived directly from the incidence relations \cite{Adamo:2017qyl}:
\be
\label{eq:ConfInv}
\mu^\da =x^{\alpha \da}\lambda_\alpha \quad \iff \quad \frac{ x_{\alpha \dal}}{x^2}\mu^\dal=\lambda_\alpha\,.
\ee
Equation \eqref{eq:ConfInv} shows that conformal inversion acts on the algebra $\mathcal{C}w_{1+\infty}$ via the automorphism \eqref{eq:ConformalInversion0}, explaining the terminology used above.


\acknowledgments

It is a pleasure to thank Tim Adamo, Roland Bittleston, Lionel Mason, John Joseph Carrasco, Monica Pate, Romain Ruzziconi, Atul Sharma, Ahmed Sheta and David Skinner for interesting conversations, and Atul Sharma for suggesting the name conformal $w$-algebra. OpenAI LLMs made significant contributions at various stages of this work. The authors take full responsibility for the final result. This work was
supported in part by NSF grant PHY-2207659, the Simons Collaboration on Celestial Holography, and
the Gordon and Betty Moore Foundation via the Black
Hole Initiative.

\begin{appendix}


\section{Details on deriving $\mathcal{C}w_{1+\infty}$}
\label{app:Details}

In this appendix, we show that the algebra $\mathcal{C}w_{1+\infty}$ is indeed fully generated by acting with $\mathfrak{so}(4,2)$ on $\mathcal{L}w_{1+\infty}$. In particular, we will see that any $C_A[a_i]$ can be represented as a sum of successive commutators involving $w^p_{\bar{m},m}\in\mathcal{L}w_{1+\infty}$ and $T^A_B\in\mathfrak{so}(4,2)$. We will use the realization of the algebras in the twistor coordinates $Z^A$ discussed in section \ref{sec:TwistorApproach}.

To this end, it will be useful to recall the definition 
\bea
\label{eq:DefineMAB}
M_{AB}[a_i]&=
a_A C_B[a_i-\delta_{iA}]
-
a_B C_A[a_i-\delta_{iB}]\\
&=\Big\{(Z^1)^{a_1}(Z^2)^{a_2}(Z^3)^{a_3}(Z^4)^{a_4},\qquad\Big\}_{AB}\,,
\eea
for $A,B\in \{1,2,3,4\}$, $(a_1,a_2,a_3,a_4)\in\mathbb{Z}^4$ with an appropriate wedge condition $a_1,a_2\geq0$ or $a_1,a_3\geq0$ (see figure \ref{fig:Wedge}) and
\be
\sum_{i=1}^4a_i=2\,.
\ee
In particular, the $\Lambda=0$ generators $w^p_{\bar{m},m}$ obeying \eqref{eq:w} can be expressed as
\be
w^p_{\bar{m},m}=\frac{1}{2}\,M_{12}[p+\bar{m}-1,p-\bar{m}-1,-p+m+2,-p-m+2]\,.
\ee
For
$\sum_{i=1}^4a_i=0\,,$ different $M_{AB}[a_i]$ obey the relation
\be
\label{eq:MABRelation}
0=(a_A+1)M_{BC}[a_i+\delta_{iB}+\delta_{iC}]+ (a_B+1)M_{CA}[a_i+\delta_{iC}+\delta_{iA}]+(a_C+1)M_{AB}[a_i+\delta_{iA}+\delta_{iB}]\,.
\ee
$M_{AB}[a_i]$ can be seen to generate arbitrary vector fields $C_A[a_i]$ as defined in \eqref{eq:CGen} through
\be
\label{eq:CfromM}
C_A[a_i]=-\frac{1}{4}\sum_{B\neq A}M_{AB}[a_i+\delta_{Bi}]\,.
\ee
The two sides of equation \eqref{eq:CfromM} differ by the following vector field, which is proportional to the homogeneity operator:
\be
\frac{1}{4}a_A (Z^1)^{a_1-\delta_{A1}}(Z^2)^{a_2-\delta_{A2}}(Z^3)^{a_3-\delta_{A3}}(Z^4)^{a_4-\delta_{A4}}\bigg(\sum_{D=1}^4 Z^D \frac{\partial}{\partial Z^D }\bigg)\,,
\ee
so that, in particular, the equality holds in the quotient algebra $\mathcal{C}w_{1+\infty}=\widetilde{\mathcal{C}}w_{1+\infty}/\mathfrak{h}$.

Hence, it only remains to show that all the $M_{AB}[a_i]$ are generated by successive commutators of $T^A_B$ with $w^p_{\bar{m},m}$.
All the generators $M_{AB}[a_i]$ with $A<B$ can be obtained using
\begin{equation}
[T^C_D,M_{AB}[a_i]]
=
a_D\,M_{AB}[a_i+\delta_{iC}-\delta_{iD}]
-\delta^C_A\,M_{DB}[a_i]
-\delta^C_B\,M_{AD}[a_i]
\end{equation}
to derive
\bea
M_{13}[a_i]
={}&
a_3\,M_{12}[a_i+\delta_{i2}-\delta_{i3}]
-
[T^2_3,M_{12}[a_i]]
\\
M_{14}[a_i]
={}&
a_4\,M_{12}[a_i+\delta_{i2}-\delta_{i4}]
-
[T^2_4,M_{12}[a_i]]
\\
M_{23}[a_i]
={}&
[T^1_3,M_{12}[a_i]]
-
a_3\,M_{12}[a_i+\delta_{i1}-\delta_{i3}]
 \\
M_{24}[a_i]
={}&
[T^1_4,M_{12}[a_i]]
-
a_4\,M_{12}[a_i+\delta_{i1}-\delta_{i4}]\\
M_{34}[a_i]
={}&[T^1_3,[T^2_4,M_{12}[a_i]]]
-a_4[T^1_3,M_{12}[a_i+\delta_{i2}-\delta_{i4}]]
-a_3[T^2_4,M_{12}[a_i+\delta_{i1}-\delta_{i3}]]
\\
&+a_3a_4\,M_{12}[a_i+\delta_{i1}+\delta_{i2}-\delta_{i3}-\delta_{i4}].
\eea
The remaining $M_{AB}[a_i]$ are generated using the manifest antisymmetry
\be
M_{AB}[a_i]=-M_{BA}[a_i]\,.
\ee

\end{appendix}
\newpage

\bibliographystyle{unsrt}
\bibliography{references}

\end{document}